# A highly scalable and energy-efficient artificial neuron using an Ovonic Threshold Switch (OTS) featuring the spike-frequency adaptation and chaotic activity

Milim Lee[1,2], Youngjo Kim[1], Seong Won Cho[1,3], Joon Young Kwak[1], Hyunsu Ju[4], Yeonjin Yi[2], Byung-ki Cheong[1], and Suyoun Lee[1,3*]

[1]*Center for Electronic Materials, Korea Institute of Science and Technology, Seoul 02792, Korea*

[2]*Institute of Physics and Applied Physics, Yonsei University, Seoul 03722, Korea*

[3]*Division of Nano & Information Technology, Korea University of Science and Technology, Daejon 34316, Korea*

[4]*Center for Opto-Electronic Materials and Devices, Korea Institute of Science and Technology, Seoul 02792, Korea*

As an essential building block for developing a large-scale brain-inspired computing system, we present a highly scalable and energy-efficient artificial neuron device composed of an Ovonic Threshold Switch (OTS) and a few passive electrical components. It shows not only the basic integrate-and-fire (I&F) function and the rate coding ability, but also the spike-frequency adaptation (SFA) property and the chaotic activity. The latter two, being the most common features found in the mammalian cortex, are particularly essential for the realization of the energy-efficient signal processing, learning, and adaptation to environments[1-3], but have been hard to achieve up to now. Furthermore, with our OTS-based neuron device employing the reservoir computing technique combined with delayed feedback dynamics, spoken-digit **recognition task has been performed with a considerable degree of recognition accuracy. From a comparison with a Mott memristor-based artificial neuron device, it is shown that the OTS-based artificial neuron is much more energy-efficient by about 100 times. These results show that our OTS-based artificial neuron device is promising for the application in the development of a large-scale brain-inspired computing system.**

The composite artificial neuron device used in this experiment is schematically presented in Figure 1(a), which is composed of an Ovonic Threshold Switch (OTS), two resistors ($R_1$ and $R_2$), and a capacitor ($C$). An OTS device has metal/amorphous chalcogenide /metal (M/$a$-ch./M) structure, which was reported to show a reversible electrical switching by S. R. Ovshinsky in 1968[4]. Due to its superior switching characteristics such as fast speed, high on/off ratio, and high current capacity, it is currently utilized as a selector device in a commercial high-density nonvolatile memory device (Optane$^{TM}$, Intel)[5,6]. The electrical switching to the ON-state (OFF-state) is considered to arise from charging (discharging) of the trap states, whose microscopic origin has not been unambiguously determined between the valence alternation pairs (VAPs) and other defects in amorphous chalcogenides[7-15]. In its OFF state, i.e., when the voltage applied to the OTS ($V_{OTS}$) is lower than a certain threshold voltage ($V_{th}$), an OTS device shows a nonlinear current-voltage ($I$-$V$) characteristic curve with high resistance ($10^7$~$10^8$ $\Omega$). When $V_{OTS}>V_{th}$, the OTS device is drastically turned on to a low resistance ($10^2$~$10^3$ $\Omega$) state. In this ON state, it shows a linear $I$-$V$ curve with low differential resistance as long as $V_{OTS}$ is higher than a holding voltage ($V_H$), below which it returns to its high resistance state completing a reversible switching (see Figure 1(d) for the characteristic $I$-$V$ curve of an OTS device).

Based on such a characteristic switching of a stand-alone OTS device, it is straightforward to understand the oscillating behavior of a composite device presented in Figure 1(a). It was first introduced as a relaxation oscillator in 1968, but its characteristics are little known[16]. Motivated by the potential of such a nonlinear oscillator that may emulate an artificial neuron[17,18], we have investigated the characteristics of the composite device focusing on its promise as a highly-scalable artificial neuron. Indeed, the scalability of an artificial neuron device is one of the most critical issues for developing a large-scale brain-inspired computing system as it was recently reported that a simple integrate-and-fire (I&F) circuit based on Si-MOSFET consumed an area of 175.3 $\mu m^2$ at 130 nm technology node[19]. We have found that our OTS-based oscillator device shows not only the basic I&F functionality and the rate coding[20], but also the spike-frequency adaptation (SFA) behavior[21] and chaotic activity[1]. The former two are the essential properties of biological neurons and have been demonstrated in a few neuromorphic devices using threshold switching devices[22-24]. To emphasize, the latter two have been hard to achieve up to now although they are the most common features found in the mammalian cortex[25] and very important for implementing the energy-efficient signal processing, learning, and adaptation to environments[1-3].

**Integrate-and-Fire (I&F) and Rate Coding**

An OTS device has a simple structure, in which an amorphous chalcogenide material is sandwiched in between electrodes as shown in Figure 1(b) and (c). In this work, a simple binary chalcogenide $Ge_{60}Se_{40}$ was used, which was reported to show robust switching behavior in our previous work[26]. Each of the bottom electrode (Mo), $Ge_{60}Se_{40}$, and the top electrode (Mo) films was deposited by RF magnetron-sputtering technique to have 100-nm thickness and patterned by a few steps of photolithography and lift-off process to have various line widths. Details of the fabrication process can be found elsewhere[27].

The characteristics of the fabricated OTS devices and the composite nonlinear oscillator devices were investigated using a measurement setup composed of a pulse generator (81110A, Agilent) and a multi-channel oscilloscope (TDS 5104, Tektronix). The basic electrical characteristics of OTS device having an electrical contact size of 5 μm are presented in the Supplementary Information S1. All the measurements were performed at room temperature at ambient conditions.

To examine the promise of the composite device as an artificial neuron, we have measured the response of the composite device to a square pulse of varying amplitude in the range of 2~10 V. Figure 2 shows the response of our composite device ($R_1$=10 kΩ, $C$=10 pF, $R_2$=100 Ω) to various stimulating voltage inputs. Figure 2(a) shows that $V_{mem}(t)$ gradually increases up to a certain level, at which a sudden drop of $V_{mem}(t)$ takes place with a concomitant rise in $I_S(t)$. This unequivocally implies that our composite device can mimic the I&F function of a biological neuron. In addition, in Figure 2(b), it is clearly shown that the average firing rate increases with increasing input voltage. To further clarify the change, the power density spectrum is plotted as a function of frequency in Figure 2(c). It is evident that the peak position of the power density shifts toward the higher frequency. This behavior is consistent with the "rate coding" of a biological neuron[20] supporting the feasibility of our composite device as an artificial neuron device.

**Spike-Frequency Adaptation (SFA)**

SFA, a gradual decrease in the spike frequency for a constant stimulus, is one of the most common properties found in excitatory neurons in the mammalian neocortex. It is known to play a critical role in energy-efficient signal processing in brains[25,28]. Looking at Figure 2(b) closely, for example, for the case of $V_{in}$=7.0 V (red), 6.8 V (orange), and 6.2 V (light blue), it can be found that the response shows the SFA behavior in some regions. Figure 3(a) presents the response of another composite device ($R_1$=10 kΩ, $C$=50 pF, $R_2$=100 Ω) with the finer time resolution. From a close look at the regions represented by red boxes, it can be noticed that the inter-spike interval (ISI) gradually increases with spikes generated recurrently. Figure 3(b) shows the magnified view of the region in the former red box in Figure 3(a), which clearly shows an elongation of $ISI_i$ with the spike count ($i$=1, 2, 3,…). In Figure 3(c), the spiking rate ($R_i$), which is defined as the reciprocal of $ISI_i$, is plotted as a function of the spike count. It is found that, in the regions designated as SFA, $R_i$ is well described by an exponential decay ($R(i)=R_0+R_1\exp(-i/\alpha)$, where $R_0$, $R_1$, and $\alpha$ are constants) represented as respective red dashed line, which is similar to the case of a biological neuron[21,29,30]. To the best of our knowledge, this is the first demonstration of the SFA behavior using the simplest artificial neuron so far unlike others using artificial neurons consisting of tens of silicon CMOS (complementary metal-oxide-semiconductor) devices[28].

As for the origin of the SFA behavior in our device, we suggest a model described in Figure 3(d)~(i), where the sequential changes in $V_{th}$ are related with those in the filling of the trap states in the OTS. In line with what was mentioned above, the electrical switching in the OTS device is described by the trap-mediated excitation of carriers followed by an avalanche. Let's assume that, in Figure 3(d), a voltage bias ($V_{bias}$) is applied to the left electrode (source) with the right one (drain) being grounded. The shaded region represents the filled trap states. When $V_{bias}= V_{th}$, the highest filled trap states touch the edge of the conduction band of a chalcogenide near the drain and the carriers in the trap states become free, turning the chalcogenide material into conductive (Figure 3(e)). As a result, the width of the OFF-region is reduced leading to the increase in the electric field in the remaining OFF-region, which accelerates the reduction of the OFF-layer width. In this manner, the OFF-region vanishes abruptly and a lot of current flow through the device instantly (Figure 3(f)). As $V_{bias}$ decreases below $V_H$, the trap states start to be evacuated below the edge of the conduction band by the emission of carriers, leading to the formation of the OFF-layer. However, the formation of the OFF-layer proceeds much slowly compared to the turning-on process due to the slow evacuation process in the weak applied field and, as a result, a hill-like distribution of the filled trap states is formed as shown in Figure 3(g). At this state, the device is in its OFF state but the effective OFF-layer is thinner than the initial OFF state resulting in the lower $V_{th}$ ($V_{th1}$) compared to the initial $V_{th}$ ($V_{th0}$). As $V_{OTS}$ increases again by charging and reaches up to $V_{th1}$, the OTS is turned on again (Figure 3(h)) but the current is lower than the first ON state because of the low bias voltage. The lower current level implies that the trap states are less filled and that, after a certain discharging period, there will be more empty trap states than in Figure 3 (g). Therefore, the OFF-layer width increases as shown in Figure 3(i) leading to the higher $V_{th}$ ($V_{th2}$) compared to $V_{th1}$. This picture explains the SFA property observed in our device in a way consistent with the behavior of $V_{mem}(t)$ and $I_S(t)$ as function of time shown in Figure 3(a) and (b).

The validity of the suggested model can be tested by investigating the spike-count dependence of the capacitance, which depends on the OFF-layer width. However, from an analysis of the time constants during the charging and discharging periods, we have found that the capacitance of the OTS device is dominated by a parasitic capacitance making an analysis of the change in capacitance hardly meaningful (see Supplementary Information S2). As an alternative to the capacitance, the delay time ($t_d$), a quiescent duration for the onset of switching after the application of a super-threshold voltage to an OTS device (see

Supplementary Information Figure S3(b)), provides information about the trap states[31,32]. For the OTS device under an electrical bias, an electron residing in trap states is supposed to move by thermally-assisted hopping process and be able to reach a critical energy level for switching after a lot of inelastic hopping processes. In this picture, $t_d$ is interpreted as a total elapsed time during the hopping processes up to the point of switching, and thus it depends on the density of empty trap states and the electric field, of which the latter is inversely proportional to the OFF-layer width. Therefore, the dependence of $t_d$ on the switching count can provide information about the change of the OFF-layer width with the switching count. To this end, we have performed a test where ten consecutive pulses are applied to an OTS device with a super-threshold voltage and $t_d$ is measured as a function of the switching count (see Supplementary Information S3). Although $t_d$ shows a large variation over different sets of measurements at each switching count because of the stochastic nature of capture/emission process, it has been clearly observed that $t_d$ tends to increase with the switching count supporting the suggested model.

As another possibility, the change in the temperature inside the OTS is conceivable in that $V_{th}$ of an OTS is known to decrease with increasing temperature[31,33-35]. In this vein, from Figure 3(a), it is noticed that the firing $V_{mem}$ has the minimum next to the big spike and gradually increases, which may be construed as due to a temperature rise from Joule heating after the first big spike followed by a slow decay. Such a change in temperature with time may ostensibly explain the observed evolution of $V_{th}$ with time. Nevertheless, the increase in the temperature is estimated to be at most 10 K using the specific heat of GeSe[36] (~20 J/g K), the density (~5 g/cm$^3$) of GeSe, the geometry of the device (area=25 um$^2$, thickness=100 nm), and the energy consumption per spike (~ 1 nJ/spike). Therefore, the increase in the temperature by Joule heating is too small to explain the observed drastic reduction in $V_{th}$ after the big spike.

**Chaotic activity**

We now focus on the analysis of the region of the irregular spiking indicated by the grey region in Figure 3(c). From measurements with tens of composite artificial neuron devices, we have found that the response can be represented by a mixture of the SFA-featured spikes and irregular spikes, where their relative weights have been found to depend on the values of $C$, $R_1$, and $R_2$ (see Supplementary Information S4). To investigate the nature of the irregular spikes, we have firstly checked if they are of chaotic nature as in the case of a biological neuron[1]. By a chaotic system, it means a system of which a few state variables are expressed in terms of coupled first-order nonlinear differential equations[37]. That is to say, it is a deterministic nonlinear dynamic system whose state variables evolve in time with a strong dependence on the initial conditions. In a single neuron and neural networks, it is known that the chaotic activity is a key ingredient in realizing adaptability and analogue computation in brains[2,38,39]. And recently, it was demonstrated through simulations that a few examples of complex problems related to the global optimization and the Traveling Salesman Problem can be solved efficiently by using an array of chaotic neuron devices[40].

For our devices, where the OTS serves as a variable resistor with resistance depending on both the bias voltage and its history, it is found that the equations describing the dynamics of our composite neuron device are consistent with the aforementioned definition of a chaotic system (see Supplementary Information S5). Apart from the mathematical definition, the experimental evidences of the chaotic nature are in the following. In Figure 4(a), a power density spectrum ($S_{V\text{mem}}$) is presented, which was obtained by Fast Fourier Transform (FFT) of $V_{mem}(t)$ of a composite neuron with $C=2$ pF, $R_1=5$ k$\Omega$, and $R_2=100$ $\Omega$ sampled at a rate of 65 MHz. It shows a broad peak around 0.3 MHz followed by a slow decay, which is described by $S_{Vmem} \sim f^{-\nu}$ with $\nu \sim 3.5$. The shape of the observed spectral dependence is quite similar to that of a squid giant axon[41,42], although the frequency range is much higher ($\sim 10^4$ times) and the slope of the decay is a little steeper in our device compared to those of a squid giant axon ($f_{max} \sim 10$ Hz, $\nu=1\sim3$).

As a typical test of a chaotic behavior of a state variable $x(t)$ of an unknown system, the time-delay embedding plot test[37] is useful, which is a plot of the trajectory of $[x(t), x(t+\Delta t),...,x(t+n\Delta t)]$ with $\Delta t$ being a delay time. For a chaotic system, the trajectory forms a confined band called as an attractor and the distance between neighboring points in the attractor increases with time due to the nonlinear nature of it. In Figure 4(b), the time-delay embedding plot of $[I_S(t), I_S(t+\Delta t), I_S(t+2\Delta t)]$ ($\Delta t=32$ nsec) is plotted, clearly showing that the trajectory resides within a band. The type of the attractor shown in Figure 4(b) is called as a strange attractor[37] which is widely observed in biological neurons[43-47]. In Figure 4(c), a trajectory of $[I_S(t), V_{mem}(t), dV_{mem}(t)/dt]$ is also plotted, which has been much studied in various biological neurons such as Onchidium pacemaker neuron[44] and giant squid axon[45,46]. It was found that $[V_{mem}(t), dV_{mem}(t)/dt]$ of biological neurons usually shows a structure of beaks and wings[45-47], which is similar to that of our device.

As a final check of the chaotic activity and the similarity between biological neurons and our device, a recurrence plot[48,49] is presented in Figure 4(d) where the value at a position in phase space of time indices $(i, j)$ is unity only if the difference between $V_{mem}(i)$ and $V_{mem}(j)$ is smaller than a certain threshold, which depends on the system under investigation, or zero otherwise. The recurrence plot obtained with the threshold of 0.7 mV, shown in Figure 4(d), is found to satisfy the three relaxed conditions of Devaney's chaos[50-52] indicating that our device is apparently a chaotic system. Furthermore, it is found that the recurrence plot is quite similar to that of the giant squid axon[52], strongly supporting the similarity of the chaotic activity of a biological neuron to that of our OTS-based composite neuron device.

Very recently, mimicking the chaotic activity of a biological neuron was demonstrated in a composite device using a Mott memristor composed of $NbO_2$[40]. Although both OTS and Mott memristor devices commonly show a threshold switching behavior, the switching mechanisms are very different from each other. In Supplementary Information S6, we present a brief comparison between those devices, where it is found that the energy efficiency of our OTS-based neuron device is remarkably improved (by two orders of magnitude) compared to the Mott memristor-based device.

**Demonstration of spoken-digit recognition**

To demonstrate the feasibility of our OTS-based neuron device as applied to real neuromorphic devices, we have implemented a benchmark test of spoke-digit recognition using a reservoir computing technique combined with delayed feedback dynamics[18,53,54]. In this technique, a single neuron device can emulate a complex neural network by taking advantage of time-multiplexing technique. For the test, we have used a subset of TI-46 database[55], which consists of 500 spoken-digit voice files taken from ten utterances of five females (10 digits×10 utterances×5 females=500 voice files).

In Figure 5(a), a schematic flowchart is shown for the spoken-digit recognition. The details of these procedures are described in Methods section. Figure 5(b) shows the result of the spoken-digit recognition test, where 450 voice files (10 digits×9 utterances×5 females) are used for training and the remaining 50 voice files are used for test. The colormap shows $<\overrightarrow{Y_{\tau i}}>_\tau$ (see Methods section) as a function of the input spoken digit ($j$) and the class index ($i$=0~9, integer), where $i$ is the digit-index for classifying an output. In order to distinguish five input voice files speaking the same digit ($j$), the $j$-values are shifted in order by 0.2. For example, all $j \in$[7.0, 7.2, 7.4, 7.6, 7.8] in Figure 5(b) indicate the same spoken-digit "seven", but are generated by five different persons. A correlation of $i \approx j$ is clearly seen rendering a high recognition accuracy (92 %). We have also performed the same test for all ten cases selecting nine utterances for training to find the average recognition accuracy of 85.4 %. These results clearly demonstrate that our OTS-based neuron device is highly promising for the application in the neuromorphic devices.

**Conclusion**

In conclusion, we have demonstrated that an artificial neuron device composed of an OTS and a few passive electrical components can mimic the behaviors of a biological neuron such as frequency-adapted I&F and chaotic activities, which are important for developing brain-inspired and low power-consuming computing devices. In addition, using our OTS-based neuron device, we have also performed a spoken-digit recognition task with a high degree of recognition accuracy. Furthermore, it is shown that the behaviors of our OTS-based neuron device are explained by a simple circuit model and can be easily controlled by adjusting the aforementioned electrical components. Finally, compared to the state of the art artificial neuron device based on Si-MOSFET, it is much superior in the scalability due to the much reduced number of necessary components. Therefore, we believe that the results presented in this work will shed a light on the way to developing a large-scale brain-inspired and low power-consuming computing systems.

**Methods**

*Fabrication and characterization of OTS devices and OTS-based neuron devices*

OTS devices were fabricated with a cross-point structure composed of metal(bottom electrode, Mo 100 nm)/Ge$_{60}$Se$_{40}$ (100 nm)/metal(top electrode, Mo 100 nm) using photolithography and lift-off techniques. Ge$_{60}$Se$_{40}$ thin films were deposited by co-sputtering technique using Ge and Ge$_{40}$Se$_{60}$ targets. The composition of Ge$_{60}$Se$_{40}$ film was analyzed by using Electron Probe Micro Analysis (EPMA) and X-ray Fluorescence (XRF) technique. The electrode (Mo) films were deposited by RF (radio-frequency) sputtering technique as well.

The composite artificial neuron devices were composed of such a fabricated OTS device, commercial resistors, and a commercial capacitor by wiring them electrically (see Figure 1(a)). The device characteristics were investigated using a measurement setup composed of a source-measurement unit (Keithley 236), a multi-channel oscilloscope (Tektronix TDS 5104), and a pulse generator (Agilent 81110A) (see Figure S1(a)).

*Demonstration of spoken-digit recognition*

Figure 5(a) shows a flowchart for the spoken-digit recognition, which consists of four steps. In the first step, an input voice waveform is divided into $N_\tau$ sections, which has been set to five in this work. Each section is analyzed to give amplitudes corresponding to frequency channels ($N_f$=83 channels in this work) releasing an intensity matrix $\overrightarrow{I_{\tau f}}$ whose size is $N_\tau \times N_f$. In the second step, a preprocessing of $\overrightarrow{I_{\tau f}}$ is performed by multiplying $\overrightarrow{I_{\tau f}}$ with a masking matrix $\overrightarrow{M_{f\theta}}$, by which each frequency components in each $\tau$-section is mapped into $N_\theta$ virtual nodes (=400 in this work) releasing $\overrightarrow{P_{\tau\theta}}$. $\overrightarrow{M_{f\theta}}$ is a $N_f \times N_\theta$ matrix whose components are randomly selected among 0, 0.41, and 0.59 with a little higher possibility of taking $0^{53}$. In the third step, each $\tau$-section in $\overrightarrow{P_{\tau\theta}}$ is sequentially concatenated releasing a time series, where each component in $\overrightarrow{P_{\tau\theta}}$ has a constant amplitude for a duration of $\theta$. Since the value of $\theta$ sets the interaction strength between virtual nodes[53], it should be optimized and is set to 400 ns in this work. The obtained time series of the preprocessed signal multiplied by an appropriate voltage gain is applied to the OTS-based neuron device by using an arbitrary function generator (Tektronix AFG-3101) and the response of our neuron device is recorded by an oscilloscope (Tektronix TDS-5104). In the final step, the time series of the recorded response is converted into a response matrix ($\overrightarrow{O_{\tau\theta}}$) with the same size as $\overrightarrow{P_{\tau\theta}}$ and is multiplied by a synaptic weight matrix $\overrightarrow{W_{\theta i}}$ giving a classifying matrix $\overrightarrow{Y_{\tau i}}$. Here, the subscript $i$ indicates a class index corresponding to digits (0~9, integer). For training a word, $\overrightarrow{Y_{\tau i}}$ is given by the following rule: 1 for the right answer and 0 otherwise for all $\tau$-sections. And using $\overrightarrow{O_{\tau\theta}}$ and $\overrightarrow{Y_{\tau i}}$, $\overrightarrow{W_{\theta i}}$ is calculated by $\overrightarrow{W_{\theta i}} = \left(\overrightarrow{O_{\tau\theta}}\right)^{\dagger} \overrightarrow{Y_{\tau i}}$, where $\left(\overrightarrow{O_{\tau\theta}}\right)^{\dagger}$ is the Moore-Penrose pseudo-inverse matrix of $\overrightarrow{O_{\tau\theta}}$. For the recognition test, $\overrightarrow{Y_{\tau i}}$ is calculated using the obtained $\overrightarrow{W_{\theta i}}$ and $\overrightarrow{O_{\tau\theta}}$ by $\overrightarrow{Y_{\tau i}} = \overrightarrow{O_{\tau\theta}} \times \overrightarrow{W_{\theta i}}$ and the maximum of the average of $\overrightarrow{Y_{\tau i}}$ over $\tau$-sections gives the guess. For this implementation, we have composed a software using LabView$^{TM}$ (National Instruments), whose processing speeds were about 3~4 words/sec for training and 20~30 words/sec for test in this work.

**Acknowledgments**

This work was supported by the Korea Institute of Science and Technology (KIST) through 2E27811. H. Ju was financially supported by National Research Foundation program, NRF-2017R1E1A1A01077484.

**Author contributions**

S.L. designed and conceived the experiments with help from M.L., Y.K., and S.W.C. M.L. and S.W.C. fabricated OTS and composite artificial neuron devices. M.L. and Y.K. performed characterization of devices. S.L. and M.L. performed the analysis of the behavior of OTS and composite neuron devices. H.J. and B.K. performed a circuit analysis about the behavior of composite neuron devices. S.L., J.Y.K., and Y.L. performed the spoken-digit recognition task. All authors discussed the data and participated in the writing of the manuscript.

**Additional information**

Extended data is available in the online version of the paper.

Reprints and permissions information is available online at http://www.nature.com/reprints.

Correspondence and requests for materials should be addressed to S.L.

**Competing interests**

The authors declare no competing interest.

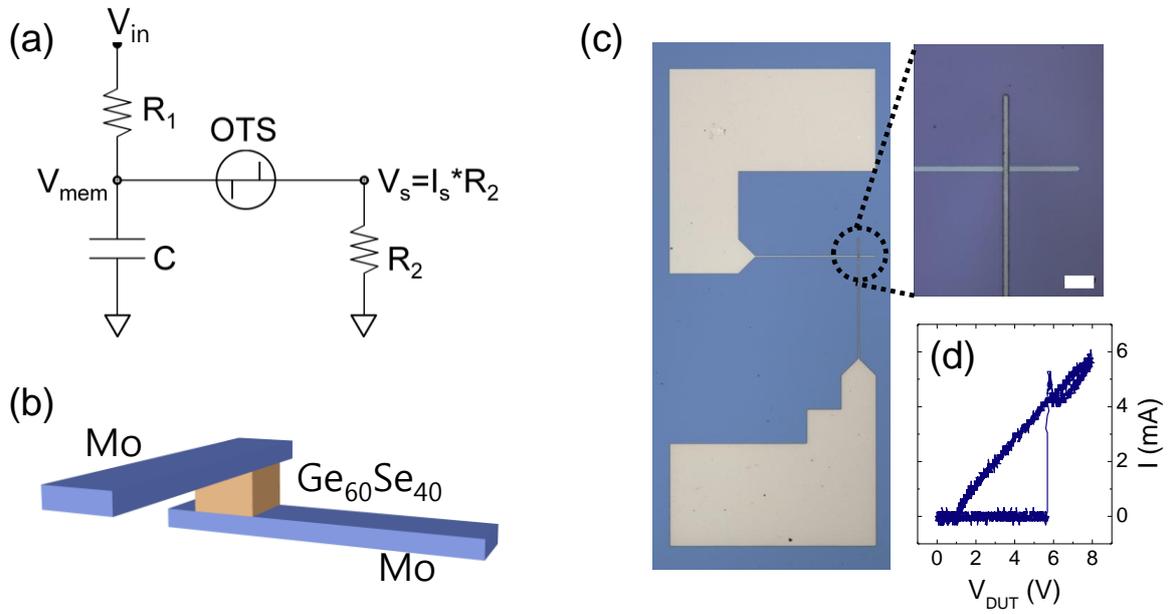

**Figure 1. An OTS device and a simple artificial neuron circuit.** (a) A simple artificial neuron circuit composed of two resistors ($R_1$ and $R_2$), one capacitor ($C$), and one OTS device. $V_{mem}$ and $I_s$ represent the membrane potential and the spike current, respectively. (b) A schematic illustration of an OTS device composed of Mo/Ge$_{60}$Se$_{40}$/Mo, (c) an optical microscope image of an OTS device which has a crosspoint sandwich structure (scale bar=20 μm). (d) The characteristic current ($I$)-voltage ($V$) curve of an OTS device.

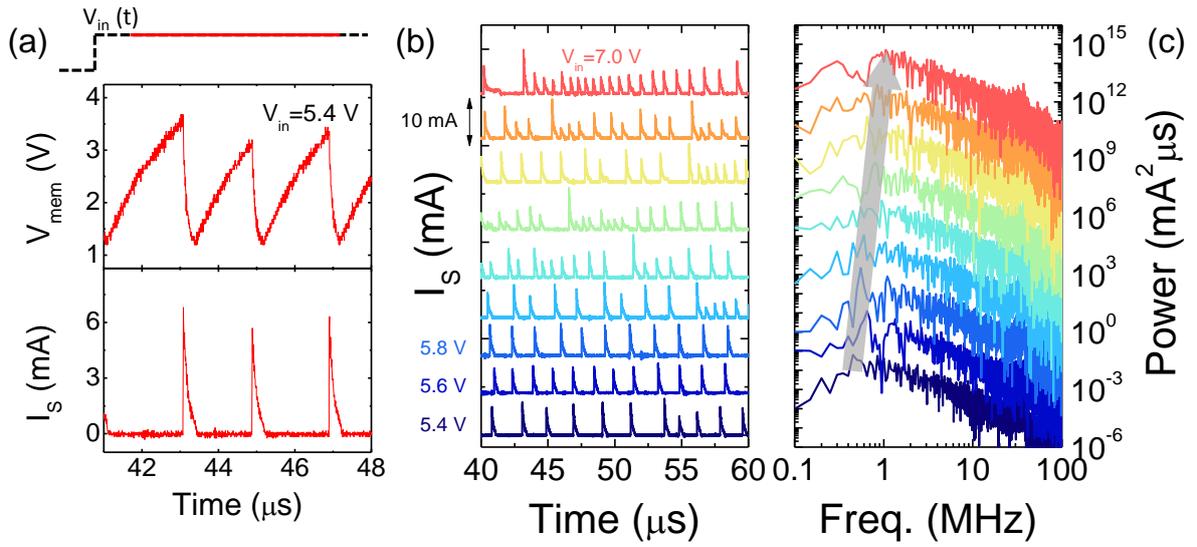

**Figure 2. Integrate-and-Fire (I&F) and rate coding of an OTS-based artificial neuron device.** (a) $V_{mem}(t)$ and $I_S(t)$ of a composite device as a response to a voltage input ($V_{in}$=5.4 V), which is a square pulse as shown on the top of the panels. The response is presented in an interval where the input is constant. (b) $I_S(t)$ as a response to various voltage inputs (5.4 ~ 7 V, 0.2 V step). The curves are shifted vertically for clarity. (c) Power density spectrum of $I_S$ for the input voltages same as (b).

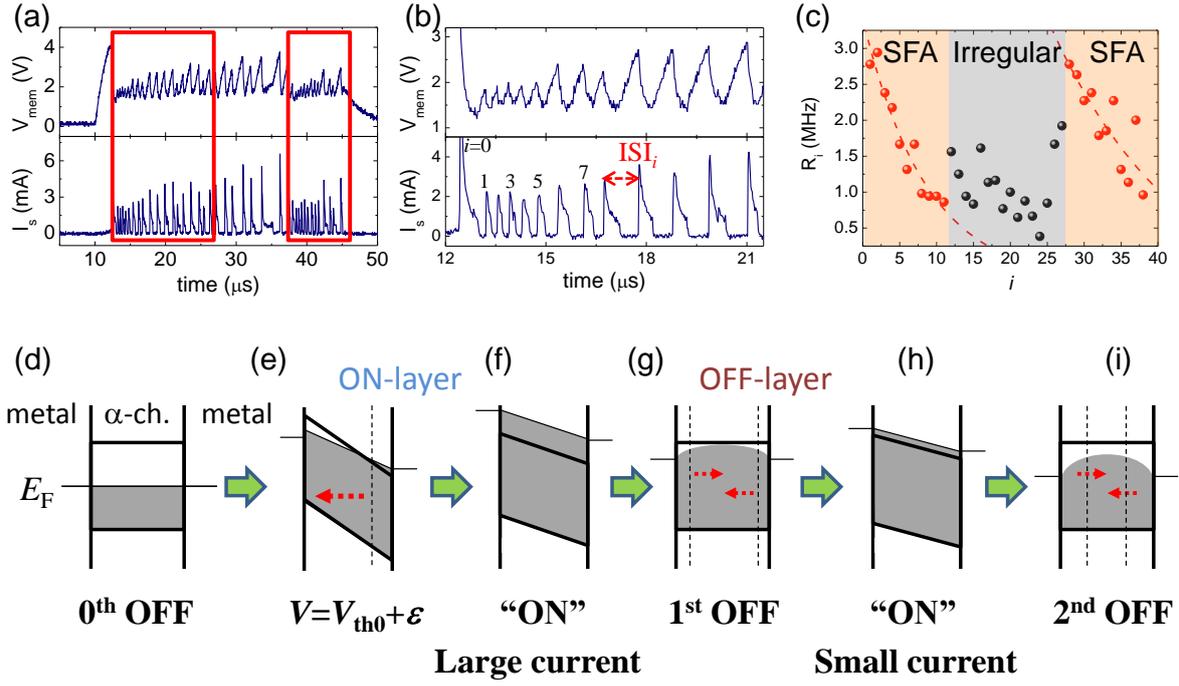

**Figure 3. Spike-frequency adaptation (SFA) behavior of an OTS-based artificial neuron device.** (a) Response of an artificial neuron circuit under a constant excitation input pulse (width=35 μs, leading edge=trailing edge=5 ns), in which $R_1$, $C$, and $R_2$ were 10 kΩ, 50 pF, and 100 Ω, respectively. (b) A magnified view in the former red box in (a) showing the SFA property, where $i$ and $ISI_i$ indicate the spike count and the inter-spike interval, respectively. (c) Spike rate ($R_i$) as a function of the spike count ($i$). Orange-colored regions represent the region showing SFA feature indicated by the red boxes in (a). The red dashed lines are fitting curves modeled by the simple exponential decay $R(i)=R_0+R_1\exp(-i/\alpha)$. (d)~(i) A sequential change in $V_{th}$ due to the change in the filling of trap states after switchings, where the shaded region represents the filled trap states. Black dashed lines represent the virtual boundary of the OFF-layer or the ON-layer, which is imagined to move along the red dotted arrows whose length represents the speed.

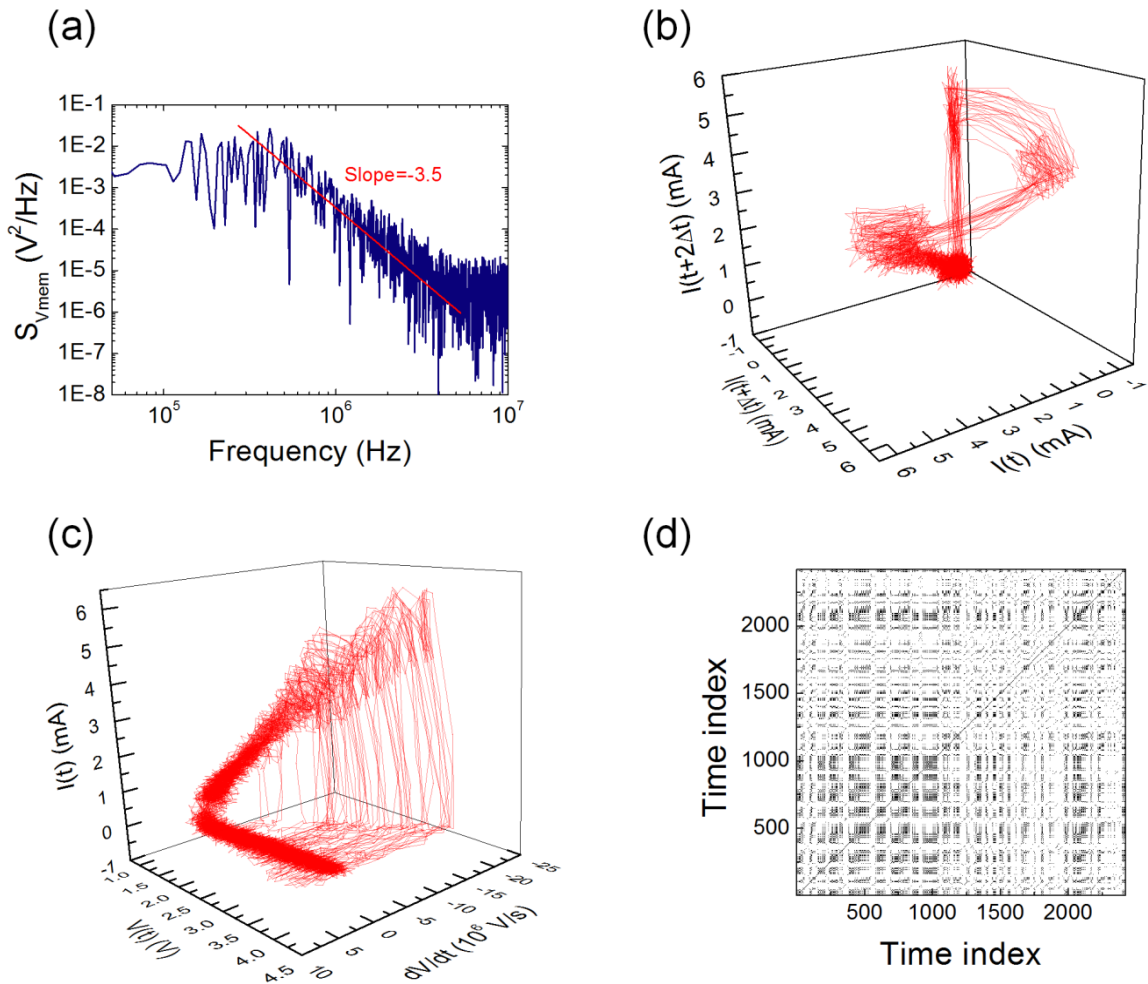

**Figure 4. Chaotic activity of the OTS-based artificial neuron.** (a) Power density spectrum of $V_{mem}$ of the artificial neuron obtained by Fast Fourier Transform (FFT), where $C$=2 pF, $R_1$=5 kΩ, $R_2$=100 Ω. (b) Trajectory of $[I_S(t), I_S(t+\Delta t), I_S(t+2\Delta t)]$ ($\Delta t$=32 nsec.) under the same $C$, $R_1$, and $R_2$ as (a). (c) Trajectory of $[I_S(t), V_{mem}(t), dV_{mem}(t)/dt]$. (d) A recurrence plot of $V_{mem}(t)$ with the threshold of 0.7 mV, by which the three relaxed conditions of Devaney's chaos are shown to be satisfied (black and white dots represent unity and zero, respectively).

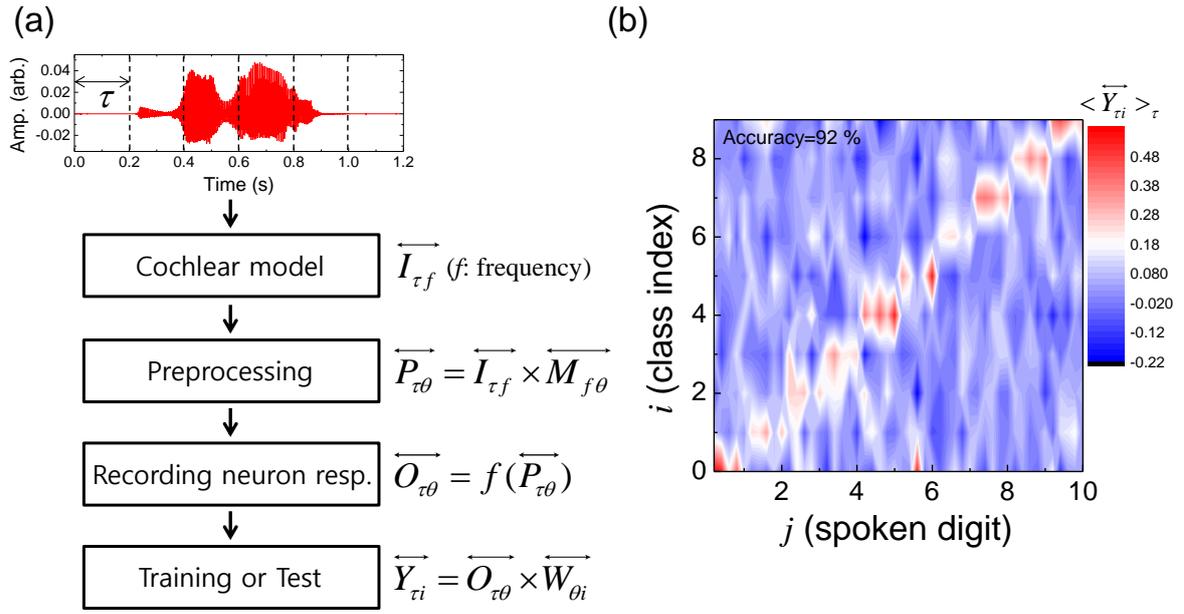

**Figure 5. Recognition of spoken-digits.** (a) Schematic flowchart for spoken-digit recognition, which is composed of four steps; (1) obtaining frequency components according to the cochlear model, (2) preprocessing procedure through which the frequency components are mapped to virtual nodes (virtual neurons), where $\overrightarrow{M_{f\theta}}$ is a masking matrix, (3) recording neuron response to the input $\overrightarrow{P_{\tau\theta}}$, and (4) training for calculating the optimum synaptic weight $\overrightarrow{W_{\theta i}}$, where $i$ is a class index (0~9, integer). As for the test in step (4), the pre-trained $\overrightarrow{W_{\theta i}}$ is used for calculating $\overrightarrow{Y_{\tau i}}$, whose average over $\tau$'s, $<\overrightarrow{Y_{\tau i}}>_\tau$, gives the guessed digit for the maximum $<\overrightarrow{Y_{\tau i}}>_\tau$. (b) A colormap of $<\overrightarrow{Y_{\tau i}}>_\tau$ as a function of the input spoken-digit ($j$=0~9, integer) and the class index ($i$), which results in the recognition accuracy of 92 %.

# References

1. CHUA, L., SBITNEV, V. & KIM, H. NEURONS ARE POISED NEAR THE EDGE OF CHAOS. *International Journal of Bifurcation and Chaos* **22**, 1250098, doi:10.1142/s0218127412500988 (2012).
2. Seifter, J. & Reggia, J. A. Lambda and the Edge of Chaos in Recurrent Neural Networks. *Artificial Life* **21**, 55-71, doi:10.1162/ARTL_a_00152 (2015).
3. Crutchfield, J. P. Between order and chaos. *Nature Physics* **8**, 17, doi:10.1038/nphys2190 (2011).
4. Ovshinsky, S. R. Reversible Electrical Switching Phenomena in Disordered Structures. *Physical Review Letters* **21**, 1450-1453 (1968).
5. DerChang, K. *et al.* in *Electron Devices Meeting (IEDM), 2009 IEEE International.* 1-4.
6. Hady, F. T., Foong, A., Veal, B. & Williams, D. Platform Storage Performance With 3D XPoint Technology. *Proceedings of the IEEE* **105**, 1822-1833, doi:10.1109/JPROC.2017.2731776 (2017).
7. Cohen, M. H., Fritzsche, H. & Ovshinsky, S. R. Simple Band Model for Amorphous Semiconducting Alloys. *Physical Review Letters* **22**, 1065-1068 (1969).
8. Anderson, P. W. Model for the Electronic Structure of Amorphous Semiconductors. *Physical Review Letters* **34**, 953-955 (1975).
9. Street, R. A. & Mott, N. F. States in the Gap in Glassy Semiconductors. *Physical Review Letters* **35**, 1293-1296 (1975).
10. Adler, D., Henisch, H. & Mott, S. The mechanism of threshold switching in amorphous alloys. *Reviews of Modern Physics* **50**, 209-220, doi:10.1103/RevModPhys.50.209 (1978).
11. Mott, N. F. Conduction in non-crystalline systems. *The Philosophical Magazine: A Journal of Theoretical Experimental and Applied Physics* **24**, 911-934, doi:10.1080/14786437108217058 (1971).
12. Ielmini, D. Threshold switching mechanism by high-field energy gain in the hopping transport of chalcogenide glasses. *Physical Review B* **78**, 035308 (2008).
13. Ielmini, D. & Zhang, Y. Analytical model for subthreshold conduction and threshold switching in chalcogenide-based memory devices. *Journal of Applied Physics* **102**, 054517, doi:10.1063/1.2773688 (2007).
14. Jennifer, L., Daniel, K., Martin, S., Matthias, W. & Christophe, L. Investigation of defect states in the amorphous phase of phase change alloys GeTe and $Ge_2Sb_2Te_5$. *physica status solidi c* **7**, 852-856, doi:doi:10.1002/pssc.200982694 (2010).
15. Caravati, S., Bernasconi, M., Kühne, T. D., Krack, M. & Parrinello, M. Coexistence of tetrahedral- and octahedral-like sites in amorphous phase change materials. *Applied Physics Letters* **91**, 171906, doi:10.1063/1.2801626 (2007).
16. Ovshinsky, S. R., Evans, E. J., Nelson, D. L. & Fritzsche, H. Radiation Hardness of Ovonic Devices. *IEEE Transactions on Nuclear Science* **15**, 311-321, doi:10.1109/TNS.1968.4325062 (1968).
17. Buzsaki, G. *Rhythms of the Brain*. (Oxford University Press, 2006).
18. Torrejon, J. *et al.* Neuromorphic computing with nanoscale spintronic oscillators. *Nature* **547**, 428, doi:10.1038/nature23011 (2017).
19. Yan, B., Cao, X. & Li, H. in *Proceedings of the 55th Annual Design Automation Conference* 1-6 (ACM, San Francisco, California, 2018).
20. Adrian, E. D. & Zotterman, Y. The impulses produced by sensory nerve-endings: Part II. The response of a Single End-Organ. *The Journal of Physiology* **61**, 151-171 (1926).

21  Wang, X.-J. Calcium Coding and Adaptive Temporal Computation in Cortical Pyramidal Neurons. *Journal of Neurophysiology* **79**, 1549-1566, doi:10.1152/jn.1998.79.3.1549 (1998).
22  Ignatov, M., Ziegler, M., Hansen, M., Petraru, A. & Kohlstedt, H. A memristive spiking neuron with firing rate coding. *Frontiers in Neuroscience* **9**, doi:10.3389/fnins.2015.00376 (2015).
23  Pickett, M. D., Medeiros-Ribeiro, G. & Williams, R. S. A scalable neuristor built with Mott memristors. *Nature Materials* **12**, 114, doi:10.1038/nmat3510, https://www.nature.com/articles/nmat3510#supplementary-information (2012).
24  Lim, H. *et al.* Relaxation oscillator-realized artificial electronic neurons, their responses, and noise. *Nanoscale* **8**, 9629-9640, doi:10.1039/C6NR01278G (2016).
25  Izhikevich, E. M. Which model to use for cortical spiking neurons? *IEEE Transactions on Neural Networks* **15**, 1063-1070, doi:10.1109/TNN.2004.832719 (2004).
26  Kim, S.-D. *et al.* Effect of Ge Concentration in $Ge_xSe_{1-x}$ Chalcogenide Glass on the Electronic Structures and the Characteristics of Ovonic Threshold Switching (OTS) Devices. *ECS Solid State Letters* **2**, Q75-Q77, doi:10.1149/2.001310ssl (2013).
27  Seo, J., Cho, S. W., Ahn, H.-W., Cheong, B.-k. & Lee, S. A study on the interface between an amorphous chalcogenide and the electrode: Effect of the electrode on the characteristics of the Ovonic Threshold Switch (OTS). *Journal of Alloys and Compounds* **691**, 880-883, doi:https://doi.org/10.1016/j.jallcom.2016.08.237 (2017).
28  Indiveri, G. *et al.* Neuromorphic Silicon Neuron Circuits. *Frontiers in Neuroscience* **5**, doi:10.3389/fnins.2011.00073 (2011).
29  Indiveri, G. Synaptic Plasticity and Spike-based Computation in VLSI Networks of Integrate-and-Fire Neurons. *Neural Information Processing - Letters and Reviews* **11**, 135-146 (2007).
30  Gabbiani, F. & Krapp, H. G. Spike-Frequency Adaptation and Intrinsic Properties of an Identified, Looming-Sensitive Neuron. *Journal of neurophysiology* **96**, 2951-2962, doi:10.1152/jn.00075.2006 (2006).
31  Lee, S. *et al.* A study on the temperature dependence of the threshold switching characteristics of $Ge_2Sb_2Te_5$. *Applied Physics Letters* **96**, 023501, doi:10.1063/1.3275756 (2010).
32  Shin, S.-Y. *et al.* The effect of doping Sb on the electronic structure and the device characteristics of Ovonic Threshold Switches based on Ge-Se. *Scientific Reports* **4**, 7099, doi:10.1038/srep07099, https://www.nature.com/articles/srep07099#supplementary-information (2014).
33  Ovshinsky, S. R. & Fritzsche, H. Amorphous semiconductors for switching, memory, and imaging applications. *IEEE Transactions on Electron Devices* **20**, 91-105, doi:10.1109/T-ED.1973.17616 (1973).
34  Shanks, R. R. Ovonic threshold switching characteristics. *Journal of Non-Crystalline Solids* **2**, 504-514, doi:https://doi.org/10.1016/0022-3093(70)90164-X (1970).
35  Lee, S. H. & Henisch, H. K. Threshold switching in chalcogenide glass films. *Applied Physics Letters* **22**, 230-231, doi:10.1063/1.1654620 (1973).
36  Borisova, Z. *Glassy Semiconductors*.  (Springer US, 2013).
37  Abarbanel, H. D. I., Brown, R., Sidorowich, J. J. & Tsimring, L. S. The analysis of observed chaotic data in physical systems. *Reviews of Modern Physics* **65**, 1331-1392 (1993).
38  Suzuki, H., Imura, J.-i., Horio, Y. & Aihara, K. Chaotic Boltzmann machines. *Scientific Reports* **3**, 1610, doi:10.1038/srep01610 (2013).
39  Kauffman, S. A. Requirements for evolvability in complex systems: Orderly dynamics and frozen components. *Physica D: Nonlinear Phenomena* **42**, 135-152, doi:https://doi.org/10.1016/0167-2789(90)90071-V (1990).

40   Kumar, S., Strachan, J. P. & Williams, R. S. Chaotic dynamics in nanoscale NbO$_2$ Mott memristors for analogue computing. *Nature* **548**, 318, doi:10.1038/nature23307, https://www.nature.com/articles/nature23307#supplementary-information (2017).

41   Conti, F., De Felice, L. J. & Wanke, E. Potassium and sodium ion current noise in the membrane of the squid giant axon. *The Journal of Physiology* **248**, 45-82, doi:10.1113/jphysiol.1975.sp010962 (1975).

42   Fishman, H. M. Relaxation Spectra of Potassium Channel Noise from Squid Axon Membranes. *Proceedings of the National Academy of Sciences* **70**, 876-879, doi:10.1073/pnas.70.3.876 (1973).

43   Hayashi, H. & Ishizuka, S. Chaotic responses of the hippocampal CA3 region to a mossy fiber stimulation in vitro. *Brain Research* **686**, 194-206, doi:https://doi.org/10.1016/0006-8993(95)00485-9 (1995).

44   Hayashi, H., Ishizuka, S. & Hirakawa, K. Instability of Harmonic Responses of Onchidium Pacemaker Neuron. *Journal of the Physical Society of Japan* **55**, 3272-3278, doi:10.1143/JPSJ.55.3272 (1986).

45   Aihara, K., Numajiri, T., Matsumoto, G. & Kotani, M. Structures of attractors in periodically forced neural oscillators. *Physics Letters A* **116**, 313-317, doi:https://doi.org/10.1016/0375-9601(86)90578-5 (1986).

46   Aihara, K., Matsumoto, G. & Ichikawa, M. An alternating periodic-chaotic sequence observed in neural oscillators. *Physics Letters A* **111**, 251-255, doi:https://doi.org/10.1016/0375-9601(85)90256-7 (1985).

47   Matsumoto, G. *et al.* Chaos and phase locking in normal squid axons. *Physics Letters A* **123**, 162-166, doi:https://doi.org/10.1016/0375-9601(87)90696-7 (1987).

48   Eckmann, J. P., Kamphorst, S. O. & Ruelle, D. Recurrence Plots of Dynamical Systems. *EPL (Europhysics Letters)* **4**, 973 (1987).

49   Marwan, N., Carmen Romano, M., Thiel, M. & Kurths, J. Recurrence plots for the analysis of complex systems. *Physics Reports* **438**, 237-329, doi:https://doi.org/10.1016/j.physrep.2006.11.001 (2007).

50   Hirata, Y. & Aihara, K. Devaney's chaos on recurrence plots. *Physical Review E* **82**, 036209 (2010).

51   Devaney, R. *An Introduction To Chaotic Dynamical Systems*.  (Westview Press, 2008).

52   Hirata, Y., Oku, M. & Aihara, K. Chaos in neurons and its application: Perspective of chaos engineering. *Chaos: An Interdisciplinary Journal of Nonlinear Science* **22**, 047511, doi:10.1063/1.4738191 (2012).

53   Appeltant, L. *et al.* Information processing using a single dynamical node as complex system. *Nature Communications* **2**, 468, doi:10.1038/ncomms1476, https://www.nature.com/articles/ncomms1476#supplementary-information (2011).

54   Larger, L. *et al.* High-Speed Photonic Reservoir Computing Using a Time-Delay-Based Architecture: Million Words per Second Classification. *Physical Review X* **7**, 011015, doi:10.1103/PhysRevX.7.011015 (2017).

55   Texas Instruments. 46-Word Speaker-Dependent Isolated Word Corpus (TI-46), NIST Speech Disk 7-1.1, httpL//catalog.ldc.upenn.edu/LDC93S9 (NIST, 1991).

# Supplementary Information

# A highly scalable and energy-efficient artificial neuron using an Ovonic Threshold Switch (OTS) featuring the spike-frequency adaptation and chaotic activity

Milim Lee[1,2], Youngjo Kim[1], Seong Won Cho[1,3], Joon Young Kwak[1], Hyunsu Ju[4], Yeonjin Yi[2], Byung-ki Cheong[1], and Suyoun Lee[1,3*]

S1.  Basic electrical characteristics of OTS devices

S2.  Calculation of the characteristic time constants during charging and discharging periods

S3.  Investigation of the change in $t_d$ with the switching count

S4.  Adjusting the behavior of the OTS-based neuron device

S5.  Validation of a nonlinear deterministic nonlinear system

S6.  A brief comparison between a Mott-memristor-based and an OTS-based artificial neuron devices.

## S1. Basic electrical characteristics of OTS devices

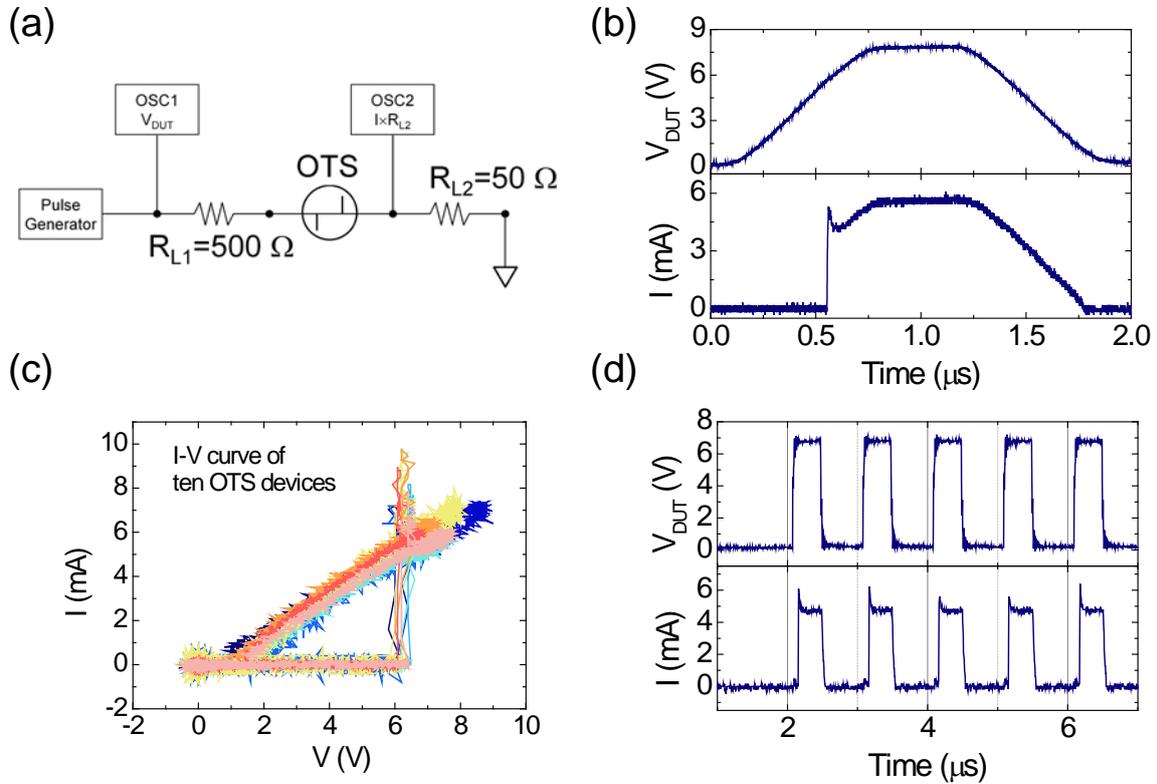

**Figure S1. Basic characteristics of an OTS device.** (a) Measurement setup for the characterization of an OTS device. (b) Response (current, bottom panel) of a typical OTS device under a trapezoidal (leading edge=500 ns, width=1.1 μs, trailing edge=500 ns) pulse (excitation, top panel). (c) Pulsed current-voltage (*I-V*) characteristic curves of ten OTS devices. (d) Response (current, bottom panel) of a typical OTS device under a train of square pulses (amplitude=6.8 V, width=300 ns, leading edge=trailing edge=5 ns, period=1 μs).

Figure S1 shows the basic characteristics of a typical OTS device. An OTS device was characterized using a measurement setup shown in Figure S1(a). As shown in Figure S1(b) and (c), the characteristic current-voltage (*I-V*) curve was measured by applying a trapezoidal pulse (leading edge=500 ns, width=1.1 μs, trailing edge=500 ns) and measuring the current to avoid a failure which would have happened due to stress by a long DC bias. It is found that our OTS device has a very high on/off ratio (~$10^5$) with the threshold voltage ($V_{th}$) and the holding voltage ($V_H$) of ~6 V and ~1 V, respectively. We also applied a pulse train to the device to find that the switching is quite reproducible as shown in Figure S1(d).

## S2. Calculation of the characteristic time constants during charging and discharging periods

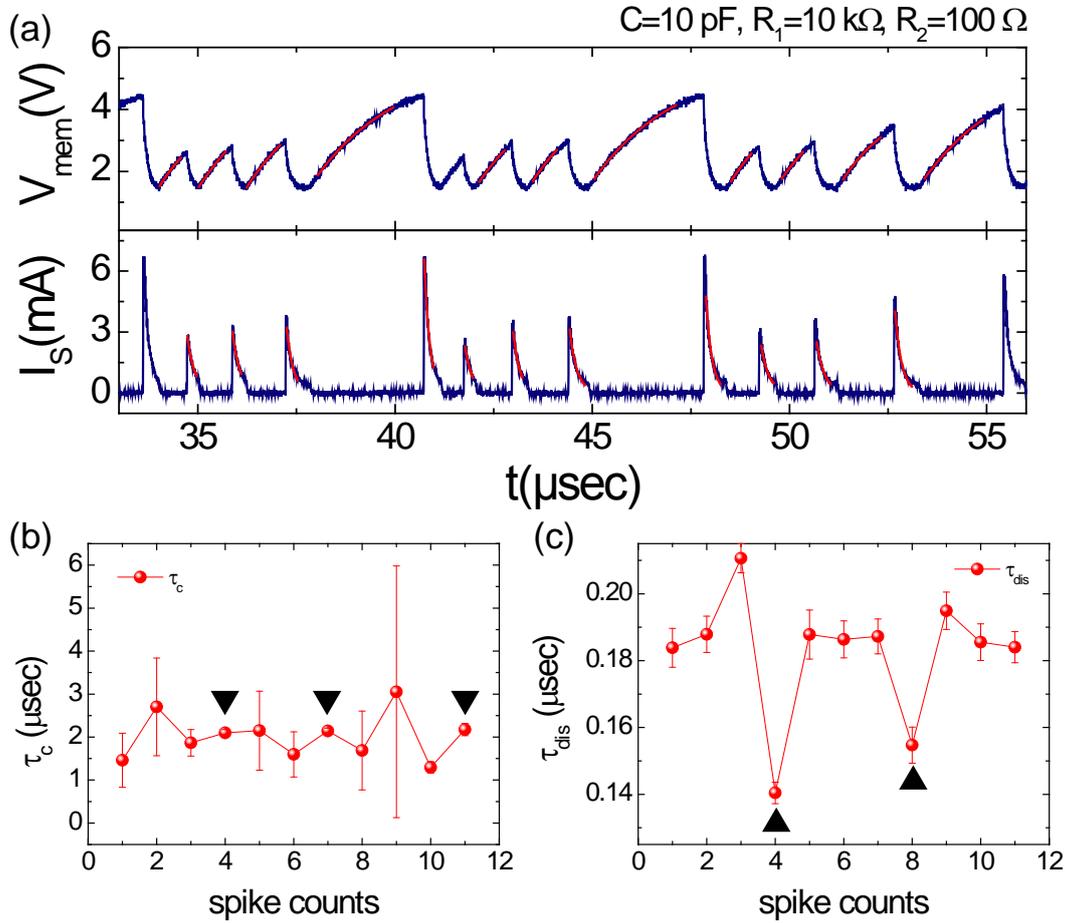

**Figure S2. Change in characteristic time constants during charging and discharging periods**, (a) $V_{mem}(t)$ (upper panel) and $I_S(t)$ (lower panel) for three regions showing typical frequency adaptation behavior. The red lines are fitted curves given by $V_{mem}(t) = V_{mem}(0)\{1-\exp[-(t-t_0)/\tau_c]\}$ and $I_S(t) = I_S(0)\exp[-(t-t_0)/\tau_{dis}]$ during the charging and discharging periods, respectively. (b) and (c) Characteristic time constants are plotted as a function of the spike count during charging periods and discharging periods, where black triangles are shown to represent the positions of big spikes.

We have investigated the change in the characteristic time constant during the charging ($\tau_c$) and discharging ($\tau_{dis}$) periods since the change in the occupation of trap states are expected to result in a change in the capacitance of the OTS device manifesting itself as a change in the time constants. The dependence of $\tau_c$ and $\tau_{dis}$ on the spike count is presented in Figure S3 in the Supplementary Information, where $\tau$'s are obtained by fitting the waveforms to $V_{mem}(t) = V_{mem}(0)\{1 - \exp[-(t-t_0)/\tau_c]\}$ and $I_S(t) = I_S(0)\exp[-(t-t_0)/\tau_{dis}]$ during the charging and discharging periods, respectively. A clear correlation between the spike count and both time constants ($\tau_c$ and $\tau_{dis}$) is not found except that $\tau_{dis}$ shows a dip at each big spike. It is found that the obtained $\tau$'s are much higher than RC delays ($R_1C$ and $R_2C$), indicating that capacitance of the OTS ($C_{OTS}$) is not negligible. Assuming an OTS as a parallel-connected capacitor and a resistor, $C_{OTS}$ is estimated to be ~ 100 pF irrespective of whether it is in ON or OFF state. Considering that independence of $C_{OTS}$ on its state despite the collapse of the band structure and the relative permittivity of GeSe ($k_{GeSe}$) of about 25[1,2], it seems to imply that $C_{OTS}$ is dominated by a parasitic capacitance.

## S3. Investigation of the change in $t_d$ with the switching count

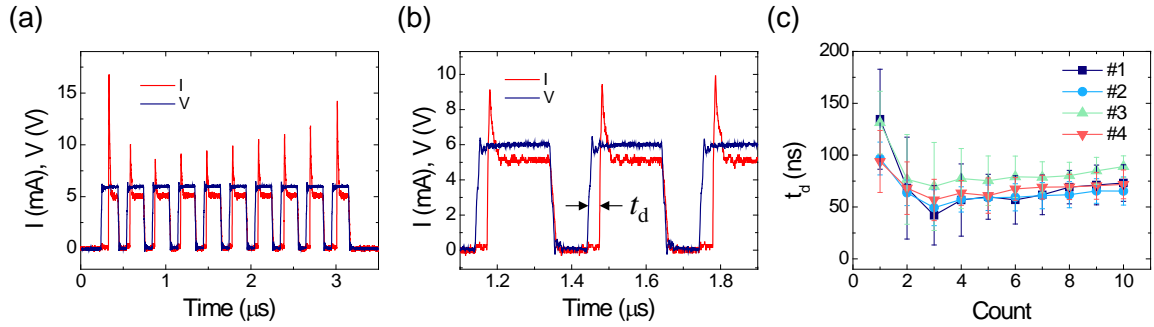

**Figure S3. Change in the delay time ($t_d$) of an OTS with the switching count.** (a) Applied (voltage) pulse train (blue line, leading+width+trailing=5+200+5 ns, period=300 ns) and the current response of an OTS device (red line), (b) a magnified view of (a) from 1.1 μs to 1.9 μs, where shows the definition of $t_d$. (c) $t_d$ as a function of the switching count for four OTS devices, where the error bar represents the standard deviation obtained from 20 repetitive measurements.

To investigate the change in $t_d$ of an OTS device with the switching count, we have performed a test where ten consecutive pulses with the over-threshold voltage are applied and $t_d$ is measured. Figure S3(a) shows a voltage pulse train (blue line, leading+width+trailing=5+200+5 ns, period=300 ns) and the current flowing through the OTS (red line). It represents the typical behavior of five OTS devices that were investigated. It is clearly shown that the peak current increases with the switching count which is similar to the SFA property of our neuron device as shown in Figure 3(b) in the main text. In Figure S3(b), we have plotted a magnified view of Figure S3(a) from 1.1 μs to 1.9 μs for the definition of $t_d$. From 20 repetitions of the same test, we have obtained the statistics of $t_d$ as a function of the switching count. Figure S3(c) shows the dependence of $t_d$ on the switching count for four OTS devices. It is commonly observed that, as switching count increases, $t_d$ gradually increases after a short initial phase of decrease. This seems to imply that the OFF-layer width increase with the switching count, supporting the model suggested in Figure 3 in the main text.

## S4.  Adjusting the behavior of the OTS-based neuron device

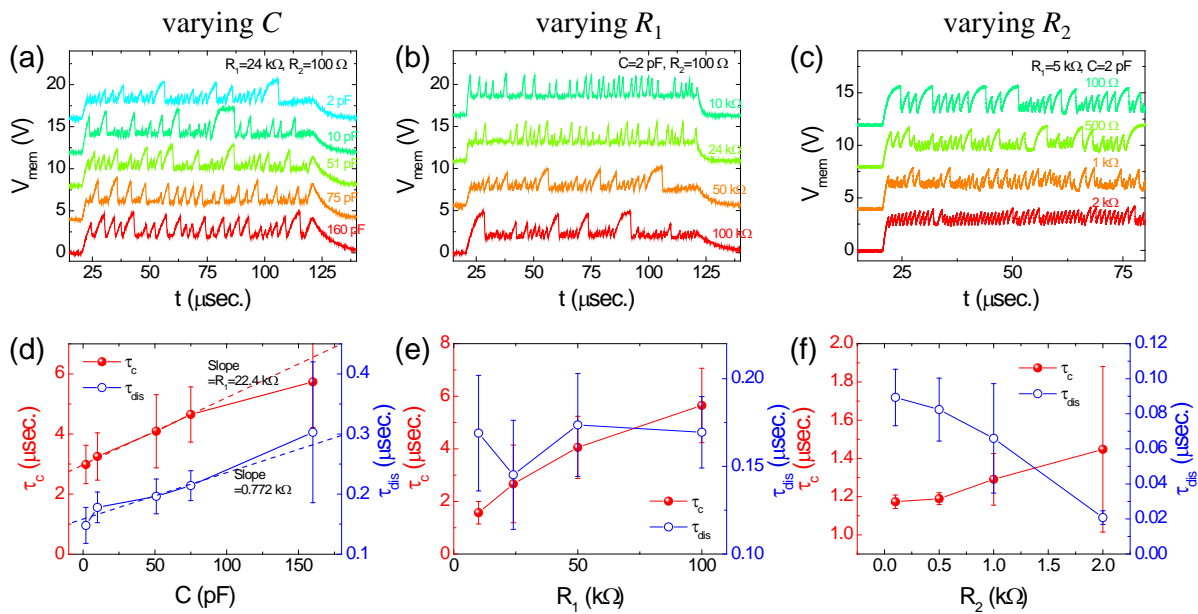

**Figure S4. Controllability of the behavior of the artificial neuron device.** (a)~(c) Change in the shape of $V_{mem}(t)$ with varying $R_1$ (a), $C$ (b), and $R_2$ (c), respectively. In respective plots, the values of the other components are fixed as indicated. The curves are vertically shifted for clarity. (d)~(f) Time constants for charging ($\tau_c$) and discharging ($\tau_{dis}$) corresponding to (a)~(c), respectively.

To assess the controllability of our artificial neuron device, we have also investigated the change in the behavior of our device with varying $C$, $R_1$, and $R_2$ to find that the shape of $V_{mem}(t)$ is quite sensitive to those values (see Figure S4). In addition, it is observed that the weights of the SFA-featured spikes and the chaotic activity-featured spikes are dependent on the values of $R_1$, $R_2$, and $C$. For more quantitative analysis, the $\tau_c$ and $\tau_{dis}$ are plotted as a function of $C$, $R_1$, and $R_2$, in Figure S4(d)~(f), respectively (for definitions of $\tau_c$ and $\tau_{dis}$, see Supplementary Information S2). Assuming the OTS as a capacitor in its OFF state, $\tau_c$ is given by $R_1(C+C_{OTS})$ during the charging period. In Figure S4(d), note that $\tau_c$ is linearly proportional to $C$ except at $C$=160 pF and the slope is 0.0224 μsec/pF=22.4 kΩ quite close to $R_1$ validating the assumed circuit model. In addition, $C_{OTS}$ is estimated to be ~133 pF from $\tau_c$ = $R_1 C_{OTS}$~3 μsec extrapolated to $C=0$. The deviation around $C=160$ pF is thought to be associated with a finite resistance of the OTS in its OFF state. The dependence of $\tau_{dis}$ on $C$ is a little more complex, leading to the slope ~ 770 Ω of the linear fitting curve, which is reasonably consistent with the sum of $R_2$ (=100 Ω) and the ON-state resistance of an OTS. In Figure S4(e), it is found that $\tau_{dis}$ is found to be nearly independent of $R_1$ as expected whereas $\tau_c$ becomes much deviated from the expected linear dependence as $R_1$ increases. It is thought to be due to the finite OFF-state resistance of the OTS similar to the explanation of the deviation at high $C$ in Figure S4(d).

The effect of $R_2$ on the oscillation behavior shown in Figure S4(c) is intriguing in that the oscillation becomes relatively uniform with reduced amplitude and shortened period as $R_2$ increases. The more intriguing finding is that $\tau_{dis}$ decreases drastically with increasing $R_2$ as shown in Figure S4(f), which is rather counter-intuitive to a simple expectation of $\tau_{dis} \sim R_2 C$. To be more realistic, the OTS is modeled as a parallel connection between a capacitor and a resistor again, where the resistance is assumed to be low (=500 Ω) in its ON state. Using a simple SPICE (Simulation Program with Integrated Circuit Emphasis) software (LTspice, Linear Technology), the transient $V_{mem}(t)$ is calculated during a discharging period (Figure S5). Interestingly, $V_{mem}(t)$ is found to show a "L"-like shape, that is to say, a sharp drop in a short time followed by a slow decay. In addition, it is noticed that the early drop in $V_{mem}(t)$ becomes sharper as $R_2$ increases, which is consistent with the observation. The "L"-like shape of $V_{mem}(t)$ is naturally explained in terms of the voltage-dividing in a series connection of the OTS and $R_2$. In detail, the characteristic discharging time from the OTS, $R_{OTS}C_{OTS}$, is approximately 66.5 nsec, much longer than $R_2 C \sim$ 4nsec with $R_2$ of 2kΩ. It indicates that discharging from the OTS mainly determines the discharging characteristic of $V_{mem}(t)$. Moreover, the larger $R_2$ pulls up the $V_{mem}(t)$ to the higher before the applied bias gets to zero.

Combining these may explain that the initial time derivative of the $V_{\text{mem}}(t)$ significantly increases to result in more rapid drop and to consequently make $\tau_{\text{dis}}$ shorter. On the other hand, the discharge current through $R_2$ accordingly slows down as the discharge time constant of $\tau_{\text{dis}} \sim R_2 C$ increases.

The above results clearly show that the behavior of our device can be easily controlled and customized by using basic circuit principles.

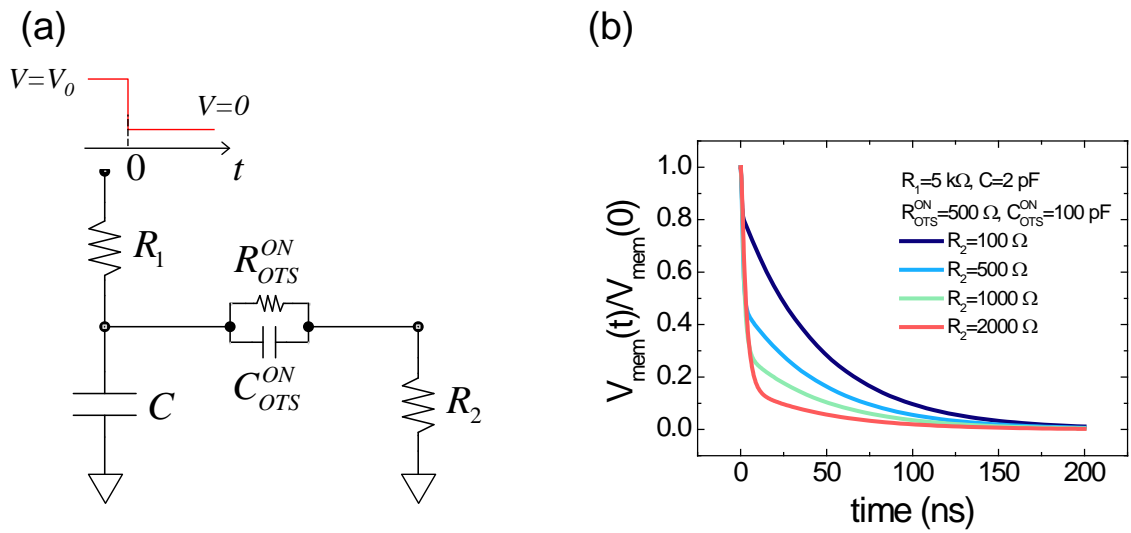

**Figure S5. Simulation of the discharging curve of an OTS-based artificial neuron.** (a) A simplified circuit model during the discharging period. After switching ON, the OTS is modeled as a parallel connection between $C_{OTS}^{ON}$ and $R_{OTS}^{ON}$ and the voltage across the OTS decreases sharply because $R_{OTS}^{ON}$ becomes low. (b) Simulation of the $V_{\text{mem}}(t)$ with varying $R_2$ by using the LTspice (Linear Technology Inc.). Values of the other components used in the simulation are also indicated.

## S5. Validation of a nonlinear deterministic nonlinear system

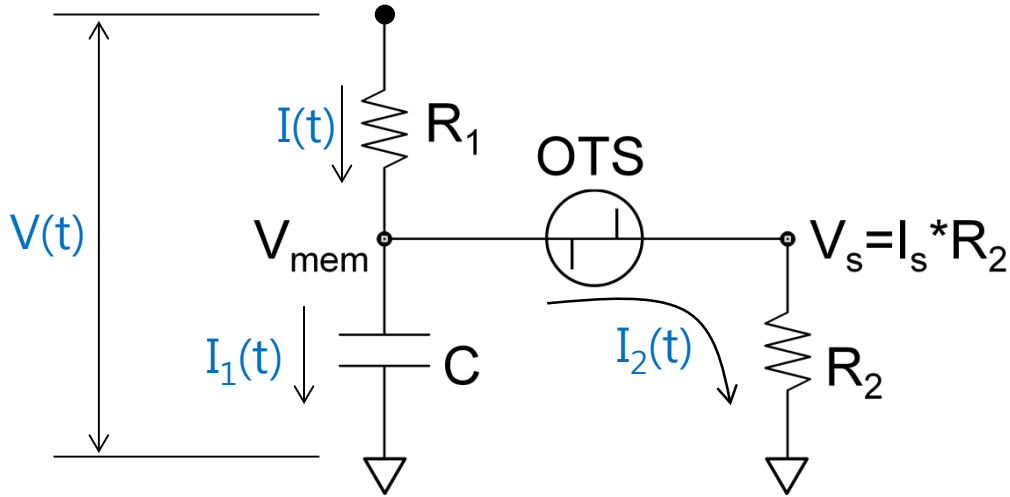

**Figure S6.** A schematic circuit of an OTS-based neuron device and the definition of state variables.

In Figure S6, we would like to calculate the dynamics of $I_S(t)$ (=$I_2(t)$ in the diagram) and $V_{mem}(t)$. If we simplify an OTS as a variable resistor ($R_{OTS}$) depending on the voltage across it ($V_{OTS}$), it is not difficult to set up equations for calculating the variables ($I(t)$, $I_1(t)$, and $I_2(t)$) using Kirchhoff's laws. However, it should be noted that an OTS is a nonlinear component, which means that $V_{OTS}$ and $R_{OTS}$ are coupled as follows. Here, $V_H$, the "holding voltage", is the voltage at which the OTS returns to its OFF state from ON state.

$$I(t) = I_1(t) + I_2(t)$$

$$V(t) = R_1 I(t) + \frac{1}{C} \int I_1(t)dt$$

$$V_{OTS}(t) = \frac{R_{OTS}}{R_{OTS} + R_2} \frac{1}{C} \int I_1(t)dt$$

$$R_{OTS}(V_{th}, V_H, t, t-\delta) = \begin{cases} R_{on} & \text{for } V_{OTS} > V_{th} \\ R_{on} & \text{for } V_H < V_{OTS} < V_{th} \text{ and } R_{OTS}(t-\delta) = R_{on} \\ R_{off} & \text{for } V_{OTS} < V_H \\ R_{off} & \text{for } V_H < V_{OTS} < V_{th} \text{ and } R_{OTS}(t-\delta) = R_{off} \end{cases}$$

From the above equations, it is clear that our OTS-based neuron device is a nonlinear deterministic system, which is a simple definition of chaos.

## S6. A brief comparison between a Mott-memristor-based and an OTS-based artificial neuron devices

|  | **Mott memristor-based[3]** | **OTS-based** |
|---|---|---|
| **Active material** | Mott insulator ($NbO_2$, $VO_2$, …) | Amorphous chalcogenide (GeSe …) |
| **Structure of a switching device** | Lance-type(presented) and crossbar (maybe possible) | Crossbar (presented) and lance-type (maybe possible) |
| **Switching mechanism** | 2-step switching<br>(1) Highly nonlinear transport relationship with temperature ("NDR-1")<br>(2) Mott metal-insulator transition ("NDR-2") | Trap-mediated excitation of carriers followed by an avalanche |
| **Integrate & Fire** | Demonstrated | Demonstrated |
| **Chaotic activity** | Demonstrated | Demonstrated |
| **Spike-frequency adaptation** | Not demonstrated | Demonstrated |
| **Energy consumption per spike per unit volume** | ~ 100 nJ/spike·$\mu m^3$ (spike current ~ 1 mA, operation voltage ~ 1 V, spike width ~ 1 $\mu s$, device size ~ 1 $\mu m^2$, thickness ~ 10 nm) | ~ 1 nJ/spike·$\mu m^3$ (spike current ~ 5 mA, operation voltage ~ 5 V, spike width ~ 0.1 $\mu s$, device size ~ 25 $\mu m^2$, thickness ~ 100 nm) |

**Table S1. A brief comparison of the Mott memristor-based and the OTS-based artificial neuron devices**

In Table S1, we have summarized the characteristics of the Mott memristor-based and the OTS-based artificial neuron device. For the Mott memristor-based neuron device, we obtained the data from a recent work[3], where $NbO_2$ was used as a Mott insulator. The structure of an artificial neuron device is same for both of the devices, where an OTS is replaced with a Mott memristor in Figure 1(a) in the main text. The parameters for calculating the energy consumption in the Mott memristor-based neuron device were obtained from Figure S7(a) in the Supplementary Information of ref. 3. From Table S1, it is found that the OTS-based neuron device is superior to the Mott memristor-based neuron device in energy efficiency. We speculate that the high energy consumption in $NbO_2$-based device is due to the operation principle, the metal-insulator transition (MIT), which requires the material to be heated up to its MIT temperature (~1070 K). In contrast, the MIT in an OTS device is of electrical origin requiring much lower energy.

**References**

1   Chandrasekhar, H. R. & Zwick, U. Raman scattering and infrared reflectivity in GeSe. *Solid State Communications* **18**, 1509-1513, doi:https://doi.org/10.1016/0038-1098(76)90381-1 (1976).

2   Siapkas, D. I., Kyriakos, D. S. & Economou, N. A. Polar optical phonons and dielectric dispersion of GeSe crystals. *Solid State Communications* **19**, 765-769, doi:https://doi.org/10.1016/0038-1098(76)90914-5 (1976).

3   Kumar, S., Strachan, J. P. & Williams, R. S. Chaotic dynamics in nanoscale NbO2 Mott memristors for analogue computing. *Nature* **548**, 318, doi:10.1038/nature23307, https://www.nature.com/articles/nature23307#supplementary-information (2017).